\documentclass{IEEEtran}
\IEEEoverridecommandlockouts
\usepackage{cite}
\usepackage{graphicx}
\usepackage{amssymb, amsmath}
\usepackage[linesnumbered,ruled,vlined]{algorithm2e}
\usepackage{bm} %
\usepackage[switch]{lineno}
\usepackage[bookmarks=false]{hyperref}
\usepackage{float}
\usepackage{booktabs}
\usepackage{subcaption}
\usepackage[T1]{fontenc}
\usepackage{newtxtext,newtxmath}
\usepackage{graphicx,cite,calc,xcolor,amssymb,amsmath,mathrsfs,dsfont,hyperref,epstopdf}
\usepackage{xcolor}
\usepackage{caption}
\usepackage[left=0.673in,right=0.667in,top=0.75in,bottom=1.1in]{geometry}

\title{QoS-Aware Energy Optimization via Cell Switching in Heterogeneous Networks}
\makeatletter
\def\@oddfoot{} 
\def\@evenfoot{} 
\makeatother
\begin{document}

 \author{\rm{Maryam Salamatmoghadasi}, \rm{Amir Mehrabian}, \rm{Halim Yanikomeroglu}, \rm{Georges Kaddoum}
 \thanks{
Maryam Salamatmoghadasi and Halim Yanikomeroglu are with the Non-Terrestrial Networks Laboratory, Department of Systems and Computer Engineering, Carleton University, Ottawa, ON K1S5B6, Canada
 (e-mails: \texttt{maryamsalamatmoghad@cmail.carleton.ca,  halim@sce.carleton.ca). }
 \\
 Amir Mehrabian and Georges Kaddoum are with the LaCIME Laboratory,
Department of Electrical Engineering, École de Technologie Supérieure,
Montreal, QC H3C 0J9, Canada (e-mails: \texttt{\{amir.mehrabian, georges.kaddoum\}@etsmtl.ca).}
}
}

\maketitle
\thispagestyle{empty}
\begin{abstract}
The growing demand for mobile data services in dense urban areas has intensified the need for energy-efficient radio access networks (RANs) in future 6G systems. In this context, one promising strategy is cell switching (CS), which dynamically deactivates underutilized small base stations (SBSs) to reduce power consumption. However, while previous research explored CS primarily based on traffic load, ensuring user quality of service (QoS) under realistic channel conditions remains a challenge. In this paper, we propose a novel optimization-driven CS framework that jointly minimizes network power consumption and guarantees user QoS by enforcing a minimum received power threshold as part of offloading decisions. In contrast to prior load-based or learning-based approaches, our method explicitly integrates channel-aware information into the CS process, thus ensuring reliable service quality for offloaded users. Furthermore, flexibility of the proposed framework enables operators to adapt system behavior between energy-saving and QoS-preserving modes by tuning a single design parameter. Simulation results demonstrate that the proposed approach achieves up to 30\% power savings as compared to baseline methods while fully maintaining QoS under diverse network conditions. Scalability and robustness of the proposed method in realistic heterogeneous networks (HetNets) further highlight its potential as a practical solution for sustainable 6G deployments.

\end{abstract}
\begin{IEEEkeywords}
  Heterogeneous networks, cell switching, energy consumption, quality of service, sustainability
\end{IEEEkeywords}
\section{Introduction}
The evolution of mobile networks from 1G to 5G has brought continuous improvements in data rates, latency, and energy efficiency, with 5G enabling advanced use cases like enhanced mobile broadband (eMBB) and ultra-reliable low-latency communication (URLLC). Building on these capabilities, 6G aims to further advance performance while prioritizing sustainability, as the environmental impact of the ICT sector increases, projected to exceed 14\% of global greenhouse gas emissions by 2040~\cite{11274940}. With 6G expected to support extremely high device densities, especially in dense urban areas, managing energy consumption is more critical than ever. Among all network components, radio access networks (RANs) are the most energy-intensive, with base stations (BSs) accounting for 60\%–80\% of total energy use~\cite{8291022,bs_power}. A promising solution is cell switching (CS), which dynamically deactivates lightly loaded or idle BSs during low-demand periods, offering substantial energy savings and supporting the sustainability goals of 6G~\cite{bs_power, 7736976, maryam,10938203}.

To date, the CS concept has been widely explored in the literature~\cite{9528008, 8735834, ELAA2022JR, EOMK2017JR, Metin_VFA_CellSwitch}, with a particular focus on reducing power consumption in cellular networks. 
In~\cite{9528008}, aiming to minimize overall network
power consumption, the authors investigated a BS switching-off strategy using traffic prediction for BSs with varying environmental characteristics in heterogeneous networks (HetNets).
A BS switching-off algorithm was also proposed in~\cite{8735834}, where the authors aimed to reduce energy consumption by selectively activating a subset of BSs.
A tiered sleep mode system that adjusts sleep depth based on device activity was proposed in~\cite{ELAA2022JR}, using decentralized control for scalability.
Furthermore, in~\cite{EOMK2017JR}, the authors explored the control data separated architecture (CDSA) and implemented a genetic algorithm to optimize energy savings in HetNets by BS deactivation through deterministic algorithms.

Another relevant investigation~\cite{Metin_VFA_CellSwitch} introduced a reinforcement learning-based CS algorithm for ultra-dense RANs, where CS was guided by traffic estimations using value function approximation (VFA). While the aforementioned approach demonstrated strong scalability and energy savings, it primarily relied on load information and did not incorporate real-time channel conditions to guarantee user quality of service (QoS).
To address this limitation, in the present study, we propose an alternative solution that jointly optimizes user association and energy consumption, with CS decisions guided by both traffic load and user channel quality. By enforcing a minimum received power threshold, our method ensures QoS for offloaded users while reducing power consumption. This channel-aware optimization framework offers a principled and flexible approach to balance energy efficiency with reliable user experience.
Major contributions of this paper can be summarized as follows:

\begin{itemize} 

\item We formulate a mixed-integer linear programming (MILP) problem that jointly optimizes BS activation and user association under capacity and QoS constraints, thus enabling energy-efficient network operation with service guarantees.

\item The CS decisions are made using real-time channel conditions to ensure that offloaded users maintain sufficient received power, thus offering a more reliable QoS than traditional load-based methods.

\item By adjusting the QoS threshold, the proposed method can flexibly operate in either a power-saving mode or QoS-preserving mode, allowing network operators to adapt based on traffic and performance priorities.

\item The algorithm is evaluated in a realistic HetNet with diverse small base station (SBS) types and power profiles. Our simulation results confirm that the algorithm achieves up to 30\% power savings while maintaining high QoS levels, thus demonstrating both practicality and robustness across varied deployment scenarios.

\end{itemize}

\section{System Model and Problem Constraints}\label{sec:system model}

\subsection{Network Model}

\begin{figure}[t]
\captionsetup{font=footnotesize}
\centerline{\includegraphics[width =8.cm ]{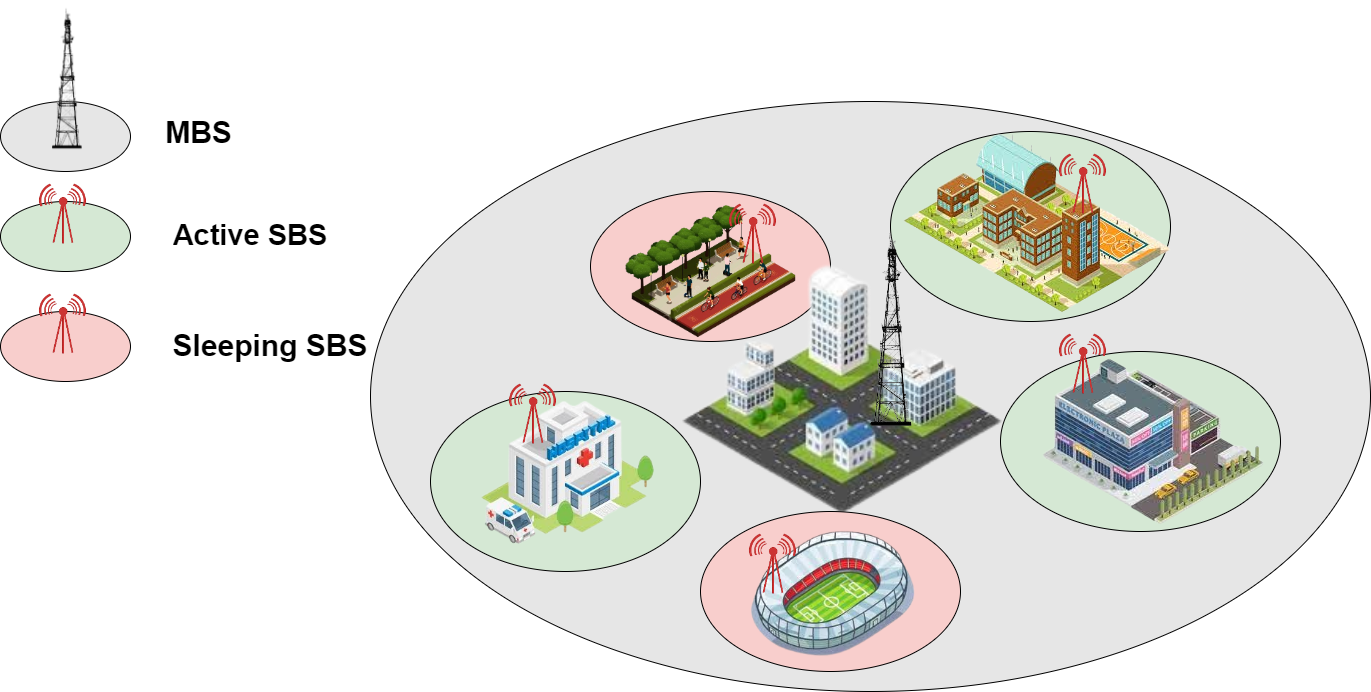}}
\caption{HetNet architecture with SBS-to-MBS offloading and energy-efficient CS.}
\label{fig1}
\vspace{-.5cm}
\end{figure}

We consider a dense HetNet composed of one macro base station (MBS) and ${s}$ SBSs, serving ${u}$ users distributed across an urban area. All BSs operate over radio frequency (RF) links and share the same spectrum. The MBS is equipped with ${N_A}$ antennas and uses Space Division Multiple Access (SDMA) to simultaneously serve multiple users via beamforming.
Each user in the network is initially associated with an SBS based on geographical proximity. However, to improve energy efficiency, the network supports dynamic offloading: specifically, during periods of low traffic or poor SBS utilization, users can be redirected to the MBS. The MBS plays a central role in maintaining network-wide coverage, managing both control signaling and fallback data services for offloaded users.
Figure~\ref{fig1} illustrates the network architecture under consideration and dynamic offloading behavior between SBSs and the MBS. The goal is to enhance sustainability by minimizing power usage while maintaining QoS across varying network loads.

\subsection{Channel Model}

To characterize the path loss between BSs and user equipment, we adopt the 3GPP-compliant channel model for urban terrestrial environments~\cite{9443997}. The total path loss is given by~\eqref{eq1}.
\begin{equation}
P_{l} = \rho^{L} P_{l}^{L} + \rho^{N} P_{l}^{N},
\label{eq1}
\end{equation}
where $\rho^{L}$ and $\rho^{N} = 1 - \rho^{L}$ are probabilities of the line of sight (LoS) and non-line of sight (NLoS), while ${P_{l}^{L}}$ and ${P_{l}^{N}}$ denote the associated losses in LoS and NLoS conditions, respectively. The LoS probability $\rho^{L}$ depends on the 2D distance $d_{2D}$ between the BS and the user and is given by~\eqref{eq2}~\cite{9443997}.
\begin{equation}
\rho ^{L}  = \left\{ {\begin{array}{*{20}c}
   {\begin{array}{*{20}c}
   1, \; \;\; \;\; \;\; \;\; \;\; \;\; \;\; \;\; \;\; \;\; \;\; \;\; \;\; \;\;\; & {d_{2D}  \le 18\,\mathrm{m},}  \\
\end{array}}  \\
   {\begin{array}{*{20}c}
   {\frac{{18}}{{d_{2D} }} + e^{( - \frac{{d_{2D} }}{{63}})(1 - \frac{{18}}{{d_{2D} }})} }, & {d_{2D}  \ge 18\,\mathrm{m},}  \\
\end{array}}  \\
\end{array}} \right.
\label{eq2}
\end{equation}

The NLoS path losse is $P_l^N = \max(P_l^L, \hat{P}_l^N)$, and the LoS path loss is computed as
\begin{equation}
P_l^L = 
\begin{cases}
P_{l1}, & 10\;\text{m} \leq d_{2D} \leq d_b, \\
P_{l2}, & d_b \leq d_{2D} \leq 5\;\text{km},
\end{cases}
\label{eq4}
\end{equation}
where 
$d_{b} = 4h'_{b} h'_{u} \frac{{f_c  \times 10^9 }}{C}$,
$h'_{b} = h_{b} - h_e$, and 
$h'_{u} = h_{u} - h_e$. We also have  
\begin{equation}
P_{l1}  = 28 + 22\log (d_{3D} ) + 20\log (f_c) + X,
\label{eq8}
\end{equation}
\begin{equation}
\begin{aligned}
P_{l2}  = 28 + 40\log (d_{3D} ) + 20\log (f_{c}) \\
-9\log (d_{b}^2  + (h_{b}  - h_{u} )^2 )+ X,
\end{aligned}
\label{eq9}
\end{equation}

\begin{equation}
\begin{aligned}
\hat P_l ^N  = 13.54 + 39.08\log (d_{3D} ) + 20\log (f_c) \\- 0.6(h_{u}  - 1.5) + X.
\end{aligned}
\label{eq11}
\end{equation}
Here, ${f_c}$ represents the carrier frequency in GHz, ${C}$ is the speed of light, and ${h_{b}}$ and ${h_{u}}$ denote the heights of the BS and user in meters, respectively, while $h_e$ represents the effective environment height, with $h_e=1$ in urban environments. In addition, ${d_{3D}}$ is the three-dimensional (3D) distance (In meters) between the BS and user. The shadow fading, represented by the log-normal random variable ${X}$, has a standard deviation of ${\sigma^{\text{L}} = 4 \text{ dB}}$ for LoS links and ${\sigma^{\text{N}} = 6 \text{ dB}}$ for NLoS links~\cite{9443997}. The received power in dBm is computed as shown in~\eqref{eq12}.
\begin{equation}
P_r = P_T + G_t + G_r - P_l,
\label{eq12}
\end{equation}
where $P_T$ is the BS transmit power, $G_t$ and $G_r$ are the transmit and receive antenna gains, and $P_l$ is the total path loss from~\eqref{eq1}.

Therefore, the channel coefficient between the $n$-th antenna of the MBS and the $i$-th offloaded user from the $j$-th SBS is denoted by $h_{i,j,n}$ and defined as
\begin{equation}
h_{i,j,n} = \frac{f_{i,j,n}}{\sqrt{P_{l,i,j}}},
\label{eq17}
\end{equation}
where $f_{i,j,n}$ represents the small-scale Rayleigh fading coefficient between the $n$-th antenna of the MBS and user, and $P_{l,i,j}$ is the large-scale path loss, computed using~\eqref{eq1}, including log-normal shadowing.

\subsection{Network Power Consumption}

We adopt the power consumption model previously defined by the energy-aware radio and network technologies (EARTH) project~\cite{earth}. The power consumption of the $j$-th SBS at any given time, denoted by $P_j$, is given by~\eqref{eq18}~\cite{7925662}.

\begin{equation}
P_j =
\begin{cases}
P_{O,j} + \eta_j \lambda_j P_{T,j}, & 0 < \lambda_j < 1, \\
P_{S,j}, & \lambda_j = 0,
\end{cases}
\label{eq18}
\end{equation}
where $P_{O,j}$ denotes the operational circuit power consumption, $\eta_j$ is the power amplifier (PA) efficiency, $\lambda_j$ represents the load factor at the $j$-th SBS, $P_{T,j}$ is the transmit power, and, finally, $P_{S,j}$ is the sleep mode circuit power. 
The total instantaneous network power consumption $P$ is then expressed as
\begin{equation}
P = P_M + \sum_{j = 1}^{s} P_j,
\label{eq19}
\end{equation}
where $P_M$ is the MBS power consumption and $s$ is the total number of SBSs. Since the MBS provides network-wide control and coverage, it remains active at all times and does not enter the sleep state. Accordingly, its power consumption follows the same model as in~\eqref{eq18}, but without the sleep mode.

\section{Problem Formulation and Solution Approach}\label{sec:problem formulation}

In this section, we present the formulation of the optimization problem for minimizing total network power consumption through selective SBS activation while maintaining user QoS.
Let $\boldsymbol{\delta} = [\delta_1, \delta_2, \dots, \delta_s]$ denote the state vector representing the activation status of each SBS, where $\delta_j \in \{0,1\}$; $\delta_j = 1$ if SBS $j$ is ON, and $\delta_j = 0$ otherwise. The MBS is always active, denoted by $\delta_M = 1$.

When an SBS is deactivated ($\delta_j = 0$), its users are redirected to the MBS. User association is determined by evaluating the received power from the MBS. A user is served by the MBS only if the received power satisfies the minimum quality threshold. 
The received signal by the $i$-th user (which initially belongs to the $j$-th SBS) from the MBS after employing linear precoding can be expressed as follows

\begin{equation}
y_{i,j} =   \mathbf{h}_{i,j}{}^H \left(\sum\limits_{i=1}^{u_{M}} \mathbf{w}_{i,j} x_{i,j}\right) + \nu_{i,j}.
\label{eq20}
\end{equation}
Here, ${\mathbf{h}_{i,j} = \left[ {h_{i,j,1} ,h_{i,j,2} ,...,h_{i,j,n} ,...,h_{i,j,N_A } } \right]^T}$ represents the RF channel vector between the MBS and the $i$-th user, offloaded from the $j$-th SBS, where $h_{i,j,n}$ is related to path loss through~\eqref{eq17}. Furthermore, ${\mathbf{w}_{i,j}} \in \mathbb{C}^{N_A }$ is the beamforming vector associated with information signal ${x_{i,j}}$, and ${\nu_{i,j}}$ is the additive white Gaussian complex noise with variance ${\frac{{\sigma ^2 }}{2}}$ on each of its real and imaginary components. Next, \(u_M\) denotes the total number of users supported by the MBS. To integrate the channel, we rewrite~\eqref{eq20} for the $p$-th user as follows

\vspace{-6pt}
\begin{equation}
\begin{array}{l}
y_{p,j} = \mathbf{h}_{p,j}{}^H \left( \mathbf{w}_{p,j} x_{p,j} + \sum\limits_{i=1,i\ne p}^{u_M}  \mathbf{w}_{i,j} x_{i,j}\right)  + \nu_{p,j}.
\end{array}
\label{eq22}
\end{equation}

For a massive multiple-input multiple-output (MIMO) system, with an increase of ${N_A}$, the L2-norms of correlated vectors grow proportionally to ${N_A}$, while the inner products of uncorrelated vectors, by assumption, grow at a lower rate~\cite{5595728}. For large ${N_A}$, only the products of identical quantities remain significant. 
Therefore, by assuming a very large number of antennas for the MBS, we can asymptotically assume that the small-scale fading vectors are orthogonal~\cite{6798744}.

\begin{equation}
\frac{{\boldsymbol{f}_{i,j}{}^H \boldsymbol{f}_{p,j} }}{{N_A}} \approx \left\{ {\begin{array}{*{20}c}
   {\begin{array}{*{20}c}
   {0,} & {p \ne i},  \\
\end{array}}  \\
   {\begin{array}{*{20}c}
   {1,} & {p = i},  \\
\end{array}}  \\
\end{array}} \right.
\label{eq23}
\end{equation}
where ${\boldsymbol{f}_{i,j} = \left[ {f_{i,j,1},...,f_{i,j,N_A } } \right]^T}$. 
Given this orthogonality assumption, we use matched filter precoding for user $p$. In this technique, precoding vector \( \mathbf{w}_{p,j} \) is set to \( \bm{f}_{p,j} / N_A \).Now, using~\eqref{eq17} and this matched filter relationship, the received signal \( y_{p,j} \) in~\eqref{eq22} is
$y_{p,j}  \approx {x_{p,j}}/{\sqrt{P_{l,p,j}}}.$
With this model, the received signal power \( P_{p,j}^r \) at user $p$ receiver is given by
$P_{p,j}^r  \approx { P_{T,M}}/{(U_M P_{l,p,j})},$
where \( P_{T,M} \) represents the maximum transmit powers for the MBS, as defined in~\eqref{eq18}. \( U_M \) denotes the total number of users that can be supported by MBS. To ensure QoS, we impose a constraint that the received power from the MBS for each user exceeds a minimum threshold, \( P_{\text{min}} \), as shown in~\eqref{eq26},
\begin{equation}
\delta _j  + (1 - \delta _j )P_{i,j}^r  \ge (1 - \delta _j )P_{\min }. 
\label{eq26}
\end{equation}
Here, if \( \delta_j = 1 \), inequality is inherently satisfied, indicating that the user’s QoS requirements are met without additional constraints. However, if \( \delta_j = 0 \) and the SBS is deactivated, then \( P_{p,j}^r \) must meet or exceed \( P_{\text{min}} \), to ensure the fulfillment of users' QoS requirements. Moreover, during the CS ON and OFF processes, additional constraints should be imposed on the loads of the MBS to ensure QoS for its users. The load on MBS at each time step is formulated as follows

\begin{equation}
\lambda _M  = \lambda _{M,0}  + \sum\limits_{j=1}^{s}{(1 - \delta _j )} \left({\sum\limits_{i=1}^{u_j}  } \lambda _{i,j}\right) \phi _{j,M}, 
\label{eq27}
\end{equation}
where \(\lambda_{M,0}\) represents the initial load on the MBS from the preceding time step. Parameter \(\lambda_{i,j}\) denotes the load contribution from the $i$-th user associated with the $j$-th SBS.  Here, \(u_j\) represents the number of users associated with the $j$-th SBS, and the total number of users across all SBSs is given by \( u=\sum_{j=1}^{s} u_j \). Furthermore, \(\phi_{j,M}\) indicates the relative capacity of the 
$j$-th SBS concerning the MBS. Specifically, \(\phi_{j,M} = {C_j}/{C_M}\), where \(C_j\) is the total capacity of the $j$-th SBS, and \(C_M\) is the total capacity of the MBS.\par
 Accordingly, we can minimize the total power consumption by solving the following optimization problem  
\vspace{-5pt} 
\begin{IEEEeqnarray*}{lcl}\label{eq:P1}
    &\underset{\mathbf{\boldsymbol{\delta}}}{\text{minimize}}\,\, & ~P(\boldsymbol{\delta} ) \,  \IEEEyesnumber \IEEEyessubnumber* \label{eq:P1_Obj}\\
    &\text{s.t.} & {\lambda _M}{\le 1,} \label{eq:P1_const1}\\
    && {{\delta _j\in \{ 0,1\}, }}  {{\;\;j=1,2,...,s}}, \label{eq:P1_const3}\\
    && {\eqref{eq26},\eqref{eq27}.} \label{eq:P1_const4} 
\end{IEEEeqnarray*}
Based on~\eqref{eq18} and~\eqref{eq19}, we can express $P(\boldsymbol{\delta} )$ as follows
\begin{equation}
\begin{aligned}
P(\boldsymbol{\delta})&={P_{O,M}  + \eta _M \lambda _M } P_{T,M}\\
 &\hspace{-2em}\;\;\;\;\;\;\;+\sum\limits_{j = 1}^s {(P_{O,j}  + \eta _j \lambda _j  P_{T,j} )\delta _j +P_{S,j}(1-\delta _j)}.
\end{aligned}
\label{eq30}
\end{equation}

Given the binary nature of state variable \(\delta_j\), the optimization problem~\eqref{eq:P1} is initially an MILP. This MILP can be efficiently solved using SCIP (Solving Constraint Integer Programs), a powerful framework capable of handling large-scale optimization problems. In our simulations, SCIP consistently produced optimal solutions within a reasonable time frame, demonstrating the practical feasibility and computational efficiency of the proposed approach even under dense network configurations.

Additionally, to simplify the model for offloading purposes, we treat all users associated with each SBS as a single group, directing them collectively to the MBS. This grouping approach assumes that users within the coverage area of a single SBS are geographically proximate, typically within a 50-meter radius. As a result, we aggregate individual loads, $\lambda_{i,j}$, into a total load value for the $j$-th SBS, denoted by $\lambda_j = \sum_{i=1}^{u_j} \lambda_{i,j}$. This aggregation allows us to simplify the problem and use a unified offloading decision variable, $\delta_j$, for all users in each SBS.

The solution approach involves solving the MILP at each time step by calculating the potential received power, forming the corresponding optimization problem, and solving it using the SCIP solver. The resulting values of $\delta_j^*$ determine which SBSs remain active and which should be offloaded to the MBS. This optimizes energy consumption while maintaining the required QoS. This process is outlined in Algorithm \ref{algo_1}, where the pseudocode details the steps for solving the MILP, including user offloading and the use of SCIP to achieve the optimal solution.
\section{Performance Evaluation}\label{sec:Performance-eva}

In this section, we evaluate effectiveness of the proposed CS strategy in a heterogeneous network with multiple types of SBSs. Simulation parameters are detailed in Table~\ref{table:nonl}. Traffic distribution across the network is modeled using a 2D Gaussian function, enabling precise control over the center (mean) and spread (variance) of the traffic load, thus representing the geographic distribution of users. Unless otherwise specified in the figure legends, all simulations assume a Gaussian model for traffic distribution. While each SBS is assigned a portion of the traffic based on its location, the initial traffic conditions for the MBS is assigned separately. This configuration enables simulating diverse traffic scenarios and effectively evaluating the performance of the proposed CS method under varying load conditions.
In the simulation setup, 49 SBSs are uniformly distributed across the area, with the MBS located at the center of the traffic load (mean). The network configuration includes the following SBS types: micro, remote radio head (RRH), pico, and femto. These SBSs--consisting of 13 micros and 12 of
each other type--are deployed across the network with nearly equal distribution. Each SBS type has distinct capacities and power consumption profiles (see Table~\ref{table: power profile}). This heterogeneous setup enables an evaluation of the performance of the proposed CS method within a realistic and diverse network environment.

\begin{algorithm}[t]
\caption{Proposed HetNet CS Optimization \label{algo_1}}
\DontPrintSemicolon  
\SetAlgoLined  
\KwData{$P_{O,j}$, $P_{O,M}$, $P_{T,j}$, $P_{T,M}$, $P_{S,j}$, $\eta _j$, $\eta _M$, $\lambda _{j}$, $\lambda _{M}$, $P_\mathrm{min}$ }
\KwResult{$\boldsymbol{\delta}^*$}

\For{all time intervals $t$}{

    \For{all SBSs with index $j$}{
        Calculate $P^r_{j,M}$\;
        
        Form the MIP optimization problem in \eqref{eq:P1}
        
        Solve the MIP using SCIP and find $\boldsymbol{\delta}^*$\;
        
        Offload all SBSs with $\delta_{j}^* = 0$ to the MBS
        
        Turn on or keep other SBSs active\;
    }
}

\label{alg1}
\end{algorithm}

\begin{table}[t]
\small
\centering 
\captionsetup{font=footnotesize}
\caption{ SIMULATION PARAMETERS}
\resizebox{.5\textwidth}{!}{
\begin{tabular}{lll}
\hline\hline 
Parameters & Notation & Typical Values\\
\hline\hline
Area size& L & $2000\; m\times 2000\; m$
\\
Carrier frequency& ${f_c}$ & 2.5 GHz 
\\
Number of SBSs& S & 49
\\
Small cell radius& R & 50 m
\\
Speed of light& ${C}$ &  299792458 m/s
\\
Transmit power of the MBS & ${P_T^M}$& 43 dBm~\cite{9583591}
\\
Antenna gain of the MBS  & ${G_T^M}$ & 8 dBi~\cite{9443997}
\\
Shadow fading standard deviation (LoS)&$\sigma ^{L}$  & 4 dB ~\cite{9443997}
\\
Shadow fading standard deviation (NLoS)& $\sigma ^{N}$ &  6 dB~\cite{9443997}
\\
Receiver sensitivity& $P_{\min }$ & -70 dBm
\\
Capacity of the SBS   & ${C_j}$ & 5
\\
Capacity of the MBS  & ${C_M}$ & 20
\\

\hline
\end{tabular}}
\label{table:nonl}
\vspace{-.3cm}
\end{table}

\subsection{Comparative Methods}
To evaluate the performance of the proposed Heterogeneous Network CS Optimization algorithm, we compare it against the following benchmark methods:

\subsubsection{All-ON}
In this method, no switching off is applied, and all SBSs remain active at all times. As a result, there are no concerns regarding QoS, since the closest SBS serves all users with the strongest received power, determined based on~\eqref{eq12}. This method serves as a baseline for evaluating power consumption, representing the scenario with no power-saving strategies.

\subsubsection{Sorting-Based CS}
Inspired by~\cite{6679195}, this method ranks SBSs based on their load factors ($\lambda$). SBSs with the lowest loads are sequentially switched off, offloading their users to the MBS, until the MBS reaches its maximum load capacity. Considering the power profiles defined in~\eqref{eq18} and summarized in Table~\ref{table: power profile}, the MBS typically consumes more power per user than SBSs. Therefore, this approach prioritizes deactivating the least loaded SBSs to minimize overall energy consumption. However, since QoS constraints are ignored, offloaded users far from the MBS may experience a reduced signal quality.

\subsubsection{CS without QoS}
In this method, we consider a variant of our proposed HetNet CS algorithm, where the QoS constraint in~\eqref{eq26} is removed. This method represents the optimization problem solved in~\cite{Metin_VFA_CellSwitch}, which optimizes the network’s power consumption by switching off SBSs and offloading traffic to the MBS. Via eliminating QoS considerations, this approach allows for a more aggressive switching-off strategy, potentially providing greater power savings at the cost of service quality. In the next section, we analyze the trade-offs between reduced power consumption and potential QoS degradation.

\begin{table}[t] 
\captionsetup{font=footnotesize}
\caption{DIFFERENT BSs POWER PROFILE~\cite{earth}}
\centering  
\resizebox{.5\textwidth}{!}{
\begin{tabular}{lllll} 
 \hline\hline 
& & &Power Consumption& \\
BS Type&Efficiency&Transmit&Operational&Sleep
\\ 
&$\eta $&$P_{T}$$[W]$&$P_{O}$$[W]$&$P_{S}$$[W]$
\\  
\hline\hline
Macro~&4.7&20&130&75
\\
RRH&2.8&20&84&56
\\
Micro&2.6&6.3&56&39 
\\
Pico&4&0.13&6.8&4.3
\\
Femto&8&0.05&4.8&2.9
\\[1ex] 
    \hline 
    \end{tabular}}
    \label{table: power profile} 
    \vspace{-0.5cm}
    \end{table}

\subsection{Simulation Results}
In this section, we evaluate the performance of the proposed Heterogeneous Network CS Optimization algorithm and compare its results with those of the benchmark methods discussed previously.

\begin{figure}[t]
    \centering
    \captionsetup{font=footnotesize}
        \includegraphics[width=.4\textwidth]{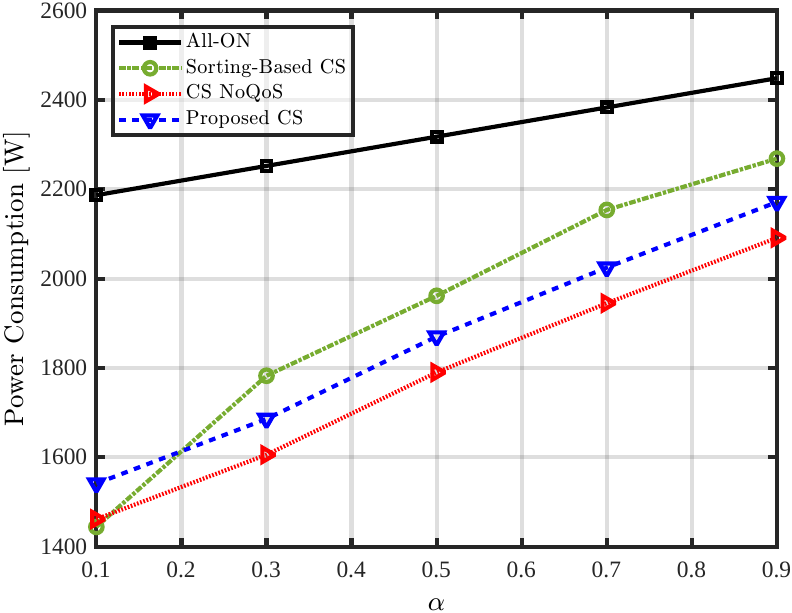}%
          \caption{Total power consumption vs. load intensity for various CS methods, with $P_\mathrm{min}=-70\; \mathrm{dBm}$.}
   \label{fig_4}
   \vspace{-.5cm}
\end{figure}

Figure~\ref{fig_4} shows the total power consumption of the network in varying load intensities, denoted by the scaling factor ($\alpha$), which reflects different levels of user demand in the Gaussian traffic distribution. The total power consumption is calculated using \eqref{eq30}. 
As illustrated in Figure~\ref{fig_4}, the All-ON method, in which all SBSs remain continuously active, exhibits the highest power consumption, which clearly indicates the inefficiency of the continuous activation of all SBSs.
As compared to this baseline, the proposed Heterogeneous Network CS Optimization achieves substantial energy savings by strategically switching off SBSs in low-demand scenarios. Specifically, at low load intensities (e.g., $\alpha=0.1$), the proposed method reduces power consumption by approximately 30\%. With an increase of load intensity, fewer SBSs can be deactivated, diminishing potential energy savings to ca. 15\% at $\alpha=0.9$.
The sorting method initially demonstrates competitive energy savings due to prioritizing the deactivation of low-load SBSs without considering QoS constraints. However, it deviates from the optimized CS approach due to the heterogeneity of the network, where different SBS types have distinct power profiles, capacities, and coverage areas. Consequently, considering the diverse
power demands of different SBS types, sorting SBSs based solely on their load becomes less effective in minimizing power consumption.
The CS without QoS allows for even more aggressive SBS deactivation as it does not impose QoS constraints on offloaded users. Thus, it achieves the lowest power consumption across all methods. However, as will be explored in subsequent figures, this advantage comes at the potential cost of reduced user service quality.

\begin{figure}[t]
    \centering
    \captionsetup{font=footnotesize}
           \includegraphics[width=.4\textwidth]{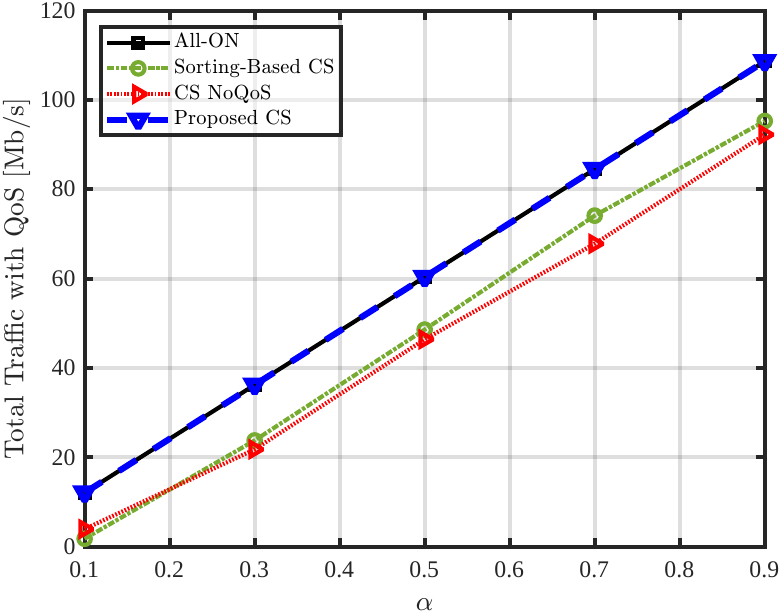}%
        \label{fig_5:b}%
    \caption{Total served traffic with QoS vs. load intensity for various CS methods, with $P_\mathrm{min}=-70\; \mathrm{dBm}$.}
    \label{fig_5}
    \vspace{-.5cm}
\end{figure}

Figure~\ref{fig_5} shows the total served traffic maintaining QoS standards ($T_{QoS}$) under varying load intensities. Here, $T_{QoS}$ is defined as the sum of the traffic handled by the MBS and all active SBSs while ensuring that each user's received power exceeds the minimum QoS threshold ($P_{\text{min}}$). Mathematically, it is given by $T_{QoS}  = T_M  +  \sum\limits_{j = 1}^s {T_j }$, where the traffic served by the $j$-th SBS, $T_j$, depends on its activation status and the received power conditions
\begin{equation}
   T_j  = 
\begin{cases}
        {C_j \lambda _j }, & {\delta _j  = 1}\;\;or \;\;({\delta _j  = 0 \;\; and \;\; P_{i,j}^r  \ge P_{\min } },) \\
        
        0, & {\delta _j  = 0\;\; and \;\;P_{i,j}^r <  P_{\min } }.
    \end{cases}
    \label{eq42}
\end{equation}

This formulation assumes that, if an SBS is active ($\delta _j  = 1$), it fully serves its assigned load, $C_j \lambda _j$. However, if the SBS is switched OFF ($\delta _j  = 0$), and the received power from the MBS at the user falls below the threshold of $P_{\min }$ (i.e., $P_{i,j}^r <  P_{\min }$), the served traffic for that SBS is set to zero, as an outage occurs due to insufficient signal quality. This metric thus evaluates how effectively the network serves traffic while maintaining QoS, particularly under various SBS configurations and offloading scenarios. As shown in Figure~\ref{fig_5}, the proposed CS consistently achieves the highest total served traffic with QoS across all load intensities, closely matching the performance of the All-ON baseline across all network loads. This outcome demonstrates the effectiveness of the Proposed CS in reducing energy consumption while fully maintaining users QoS.
By contrast, the CS NoQoS and sorting-based CS methods result in lower served traffic levels. Since both methods do not enforce the QoS constraint during SBS offloading, some users may experience degraded received power, leading to service outages and reducing the total handled traffic. 

\begin{figure}[t]
    \centering
    \captionsetup{font=footnotesize}
    
        \includegraphics[width=.4\textwidth]{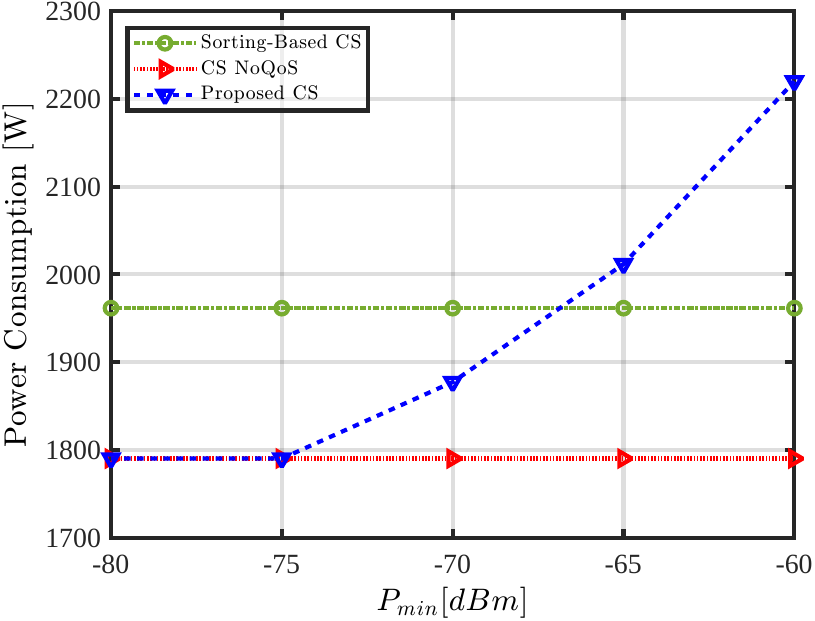}%
        \label{fig_6:b}%
    \caption{Total power consumption vs. $P_\mathrm{min}$ for the proposed CS method and benchmark algorithms, with $\alpha = 0.5$.}
    \label{fig_6}
    \vspace{-.5cm}
\end{figure}

Figure~\ref{fig_6} shows the total power consumption of the network as a function of $P_\mathrm{min}$, with simulations conducted for a fixed load intensity ($\alpha = 0.5$). As can be seen in the results, only the proposed CS approach is affected by variations in $P_\mathrm{min}$, while the other methods maintain constant power consumption regardless of $P_\mathrm{min}$.
At lower threshold values ($P_\mathrm{min} \leq -75;\mathrm{dBm}$), the proposed CS behaves similarly to CS NoQoS, as the QoS requirement is easily met and most users already meet the received power condition. As $P_\mathrm{min}$ increases and becomes more stringent, the gap between the two approaches gets larger.
The proposed CS responds to stricter QoS demands by activating more SBSs to maintain user service quality, which results in a rise in power consumption. By contrast, provided the MBS capacity
constraint is respected, CS NoQoS continues to minimize energy use by switching off as many SBSs as possible, without accounting for users’ QoS requirements.
This behavior highlights the trade-off between energy efficiency and QoS assurance inherent in the proposed method. By appropriately selecting $P_\mathrm{min}$, network operators can tune the balance between minimizing power consumption and preserving service quality.
Moreover, the results emphasize that the proposed CS framework offers considerable flexibility: setting a very low $P_\mathrm{min}$ effectively transforms the proposed CS into the CS NoQoS scheme, which demonstrates its adaptability to different network priorities.

\section{Conclusion}\label{sec:con}
 In this study, we presented a novel CS optimization framework for energy-efficient operation in heterogeneous terrestrial networks. The proposed method jointly considers user association and SBS activation, using both traffic load and real-time channel conditions to ensure user QoS while minimizing power consumption. Our simulation results demonstrated that the proposed approach achieves up to 30\% energy savings as compared to baseline methods, with consistent QoS guarantees. Moreover, by adjusting the QoS threshold, the framework can flexibly adapt to different network priorities—ranging from aggressive energy conservation to strict QoS assurance—which makes it a scalable and practical solution for next-generation mobile networks.

\section*{Acknowledgment}

This study was supported in part by the Connected and Autonomous Vehicles (TrustCAV) CREATE Program funded by the Natural Sciences and Engineering Research Council of Canada (NSERC).

\ifCLASSOPTIONcaptionsoff
  \newpage
\fi
\small

\end{document}